\documentclass[%
reprint,
superscriptaddress,
 amsmath,amssymb,
prb,
]{revtex4-2}

\usepackage{graphicx}
\usepackage{dcolumn}
\usepackage{bm}
\usepackage{hyperref}

\begin{document}

\preprint{APS/123-QED}

\title{Gate voltage dependent Rashba spin splitting in hole transverse magnetic focussing}
\author{M. J. Rendell}
    \thanks{M. J. Rendell and S. D. Liles contributed equally to this work}
\author{S. D. Liles}
\author{A. Srinivasan}
\affiliation{
School of Physics, University of New South Wales, Sydney, NSW 2052, Australia
}
\author{O. Klochan}
\affiliation{
University of New South Wales Canberra, Canberra, ACT 2600, Australia
}
\affiliation{
School of Physics, University of New South Wales, Sydney, NSW 2052, Australia
}
\author{I. Farrer}
\affiliation{
Department of Electronic and Electrical Engineering, University of Sheffield, Sheffield, S1 3JD, UK
}
\affiliation{
Cavendish Laboratory, University of Cambridge, Cambridge, CB3 0HE, UK
}
\author{D. A. Ritchie}
\affiliation{
Cavendish Laboratory, University of Cambridge, Cambridge, CB3 0HE, UK
}
\author{A. R. Hamilton}
\email{alex.hamilton@unsw.edu.au}
\affiliation{
School of Physics, University of New South Wales, Sydney, NSW 2052, Australia
}
\date{\today}

\begin{abstract}
Magnetic focussing of charge carriers in two-dimensional systems provides a solid state version of a mass spectrometer. In the presence of a spin-orbit interaction, the first focussing peak splits into two spin dependent peaks, allowing focussing to be used to measure spin polarisation and the strength of the spin-orbit interaction. In hole systems, the $k^3$ dependence of the Rashba spin-orbit term allows the spatial separation of spins to be changed in-situ using a voltage applied to an overall top gate. Here we demonstrate that this can be used to control the splitting of the magnetic focussing peaks. Additionally, we compare the focussing peak splitting to that predicted by Shubnikov-de Haas oscillations and $k\cdot p$ bandstructure calculations. We find that the focussing peak splitting is consistently larger than expected, suggesting further work is needed on understanding spin dependent magnetic focussing.
\end{abstract}

\pacs{Valid PACS appear here}
\maketitle


\section{\label{sec:Intro}Introduction}
Magnetic focussing is the solid state realisation of a mass spectrometer. Originally proposed as a way to measure the Fermi surface in metals,\cite{sharvin_possible_1965, tsoi_focusing_1974} it has subsequently been used to probe band structures in graphene,\cite{taychatanapat_electrically_2013} spatially separate spin states,\cite{rokhinson_spin_2004, heremans_spin-dependent_2007, lo_controlled_2017}, extract electron-electron scattering lengths,\cite{gupta_precision_2021} and measure spin polarisation.\cite{potok_detecting_2002,folk_gate-controlled_2003,rokhinson_spin_2004,rokhinson_spontaneous_2006,chen_all-electrical_2012}

In 2D hole systems in GaAs, magnetic focussing has been used to measure scattering\cite{heremans_observation_1992, rendell_transverse_2015} and demonstrate spatial separation of spin.\cite{rokhinson_spin_2004} By spatially separating the spin states, the spin polarisation in the 2D hole system could be measured using the amplitude of magnetic focussing peaks.\cite{rokhinson_spin_2004} This technique was subsequently used to study the transmission of spin by a quantum point contact (QPC) in a focussing setup.\cite{rokhinson_spontaneous_2006, chesi_anomalous_2011}

While previous work on hole magnetic focussing has concentrated on the amplitude of the magnetic focussing peaks, the spacing of the focussing peaks can also be used to obtain information about the spin-orbit interaction of holes. In this work, we demonstrate control over focussing peak splitting in-situ using a voltage applied to an overall top gate. We then compare the magnitude of this splitting to predicted values from bandstructure calculations and Shubnikov-de Haas oscillations, and find the splitting is consistently larger than expected. This result suggests that focussing peak splitting alone is not a good measure of the Rashba spin-orbit interaction in 2D hole systems.
\section{\label{sec:HolesVsElectrons}Bandstructure of two-dimensional hole systems}
\begin{figure*}
    \centering
    \includegraphics[width=\textwidth]{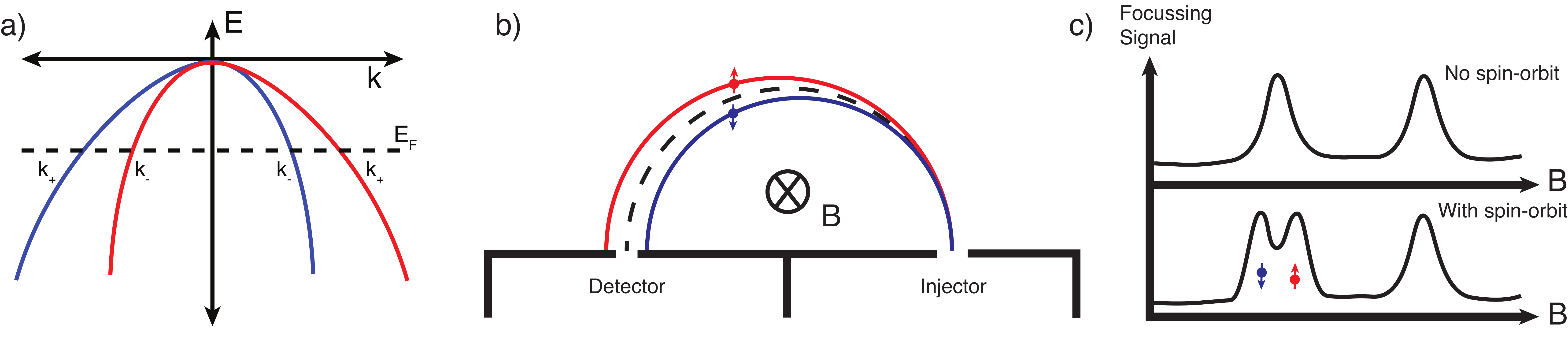}
    \caption{\textbf{a)} 2D band dispersion of holes with a Rashba SOI. The $k^3$ Rashba term splits the first heavy hole subband into two spin chiralities (red and blue curves). This results in two spin dependent values of momentum ($k_+$ and $k_-$) at the Fermi energy (dashed horizontal line). \textbf{b)} Schematic of magnetic focussing with a spin-orbit interaction. The black dashed line represents the classical focussing trajectory in the absence of a spin-orbit interaction. When a spin-orbit interaction is present, two spin dependent trajectories are created (blue and red paths). \textbf{c)} Focussing peaks with and without a spin-orbit interaction. In the presence of a spin-orbit interaction the first focussing peak splits into two spin-dependent peaks.}
    \label{fig:SpinResolvedFocussing}
\end{figure*}
2D hole systems are fundamentally different to equivalent electron systems. One key difference arises from the form of the Rashba spin-orbit interaction, and this difference can have a dramatic impact on spin resolved focussing in hole systems.

In hole systems, the Rashba spin-orbit interaction has a $k^3$ dependence. In GaAs, the subband dispersion for 2D holes with a Rashba spin-orbit interaction is given by\cite{winkler_spin-orbit_2003}
\begin{equation}\label{eq:HoleRashba}
    \mathcal{E}_{h} = \frac{\hbar^2 k^2}{2m^*} \pm \beta \frac{E_z}{\Delta_{HH-LH}} k^3
\end{equation}
where $\beta$ is a constant, $E_z$ is the electric field in the out-of-plane direction and $\Delta_{HH-LH}$ is the splitting between the heavy hole (HH) and light hole (LH) subbands. Figure \ref{fig:SpinResolvedFocussing}a) shows the resulting HH subband dispersion for a 2D hole system with Rashba SOI. The Rashba SOI creates two spin dependent $k$ values at the Fermi energy (horizontal dashed line), resulting in a spatial separation of spin in the 2D region. In addition, the magnitude of the Rashba spin-orbit interaction can be changed using a voltage applied to an overall top gate ($V_{TG}$) on the focussing sample. $V_{TG}$ will change $E_z$ and $k$, while only having a small effect on $\Delta_{HH-LH}$. Therefore it is possible to change the magnitude of the Rashba spin-orbit interaction and hence the focussing peak splitting in-situ by changing $V_{TG}$. The ability to tune the focussing peak splitting using $V_{TG}$ serves as the focus of this paper.
\section{\label{sec:Focussing}Sample and magnetic focussing setup}
\begin{figure*}
	\includegraphics[width=\textwidth]{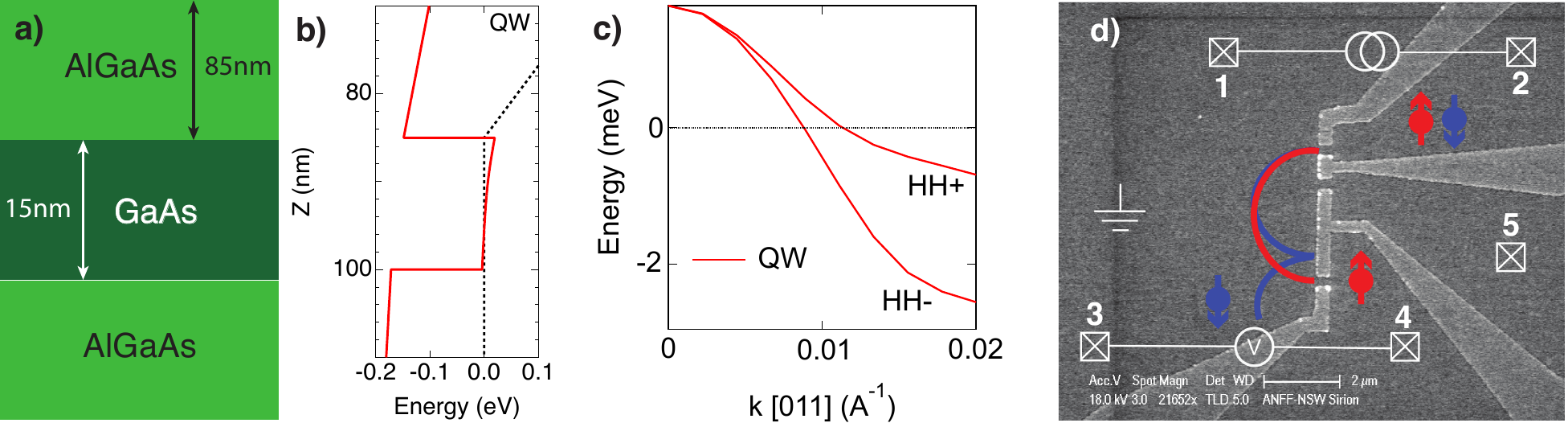}
    \caption{\textbf{a)} Schematic of the QW wafer - a 15nm $GaAs$ quantum well in a $GaAs/Al_{0.33}Ga_{0.67}As$ heterostructure (wafer W713). \textbf{b)} The square well confining potential created by the heterostructure in \textbf{a)} from a Schrodinger-Poisson calculation using Nextnano. \textbf{c)} Band dispersion of the spin-split HH subbands for the QW, calculated using a 6x6 k.p solver (Nextnano). \textbf{d)} SEM image of the magnetic focussing lithography with an overlaid electrical setup. Red and blue semicircles show the trajectories of different spins in the presence of a spin-orbit interaction. Three different focussing diameters ($d_{Focus}$) are available through use of different QPC combinations - 800nm, 2300nm and 3100nm.}
    \label{fig:FocussingDiagram}
\end{figure*}
Figure \ref{fig:SpinResolvedFocussing}b) shows a schematic of a hole transverse magnetic focussing device. A constant current is applied through the injector and a perpendicular out-of-plane magnetic field causes holes to form cyclotron orbits. The detector voltage is measured as a function of magnetic field strength, and peaks in the focussing signal occur when the focussing diameter matches the distance between injector and detector. In the absence of a spin-orbit interaction, the holes follow a single trajectory (black dashed line). In the presence of a spin-orbit interaction, the focussing trajectory becomes spin dependent (red and blue lines).\cite{rokhinson_spin_2004, zulicke_magnetic_2007, bladwell_magnetic_2015} Figure \ref{fig:SpinResolvedFocussing}c) compares the focussing signal with and without a spin-orbit interaction. When there is a spin-orbit interaction, the first focussing peak splits into two spin-dependent peaks. This splitting of the first peak has been used to measure the strength of the spin-orbit interaction.\cite{rokhinson_spin_2004} The second focussing peak does not split due to boundary reflection causing a refocusing of the spin dependent trajectories.\cite{usaj_transverse_2004, lee_influence_2021}

To investigate the $V_{TG}$ dependence of hole focussing, a magnetic focussing sample was fabricated on an undoped, accumulation mode $GaAs/Al_{0.33}Ga_{0.67}As$ heterostructure. The heterostructure is shown in Figure \ref{fig:FocussingDiagram}a), and consists of a 15nm $GaAs$ quantum well (QW) 85nm below the wafer surface (Figure \ref{fig:FocussingDiagram}b). In order to accumulate a 2D hole gas (2DHG) in the well, 30nm of $Al_2O_3$ is deposited using atomic layer deposition to act as a gate oxide, followed by an overall Ti/Au top gate.

The asymmetric potential confining the 2D holes leads to a Rashba splitting of the valence band (Figure \ref{fig:FocussingDiagram}c) creating a difference in momentum between the spin chiralities which is detected using magnetic focussing.\cite{rokhinson_spin_2004} We use Nextnano\footnote{\url{http://www.nextnano.de/}} to calculate the subband energies and $E(k)$ dispersions of the 2DHS. The Nextnano calculations employ a combination of a Schrodinger-Poisson solver and a $6\times6$ $k\cdot p$ calculation for our sample heterostructure. The calculations include the Rashba SOI term but do not include contributions from Dresselhaus SOI. Figure \ref{fig:FocussingDiagram}c) shows the calculated HH1 subband dispersion for our sample, with a different $k_F$ visible at the Fermi energy (dashed horizontal line) for the two different spin states. The different $k_F$ results in a different cyclotron radius for each spin (HH+ and HH-), causing a spatial separation of spin in the 2D region in focussing.\cite{rokhinson_spin_2004}

A scanning electron microscope (SEM) image of the device before the top gate is deposited is shown in Figure \ref{fig:FocussingDiagram}d). Metal split gates are used to define 1D quantum point contacts (QPCs) used to inject and detect the focussed beam. The QPCs are symmetrically biased (minimal $\Delta V_{SG}$) to the $G=2e^2/h$ conductance plateau to allow for the injection and detection of both spin chiralities.\cite{rokhinson_spontaneous_2006} Symmetrically biasing the QPCs at $G=2e^2/h$ also avoids the complex structure of the first hole subband\cite{hudson_new_2021} and additional spin dynamics created by the QPC.\cite{lo_controlled_2017} A constant current (5nA) is injected through one QPC (contacts 1 and 2) using a lockin amplifier, while the voltage buildup is measured across a second QPC (contacts 3 and 4), allowing a four-terminal focussing resistance to be measured ($R_{Focus} = V_{34}/I_{12}$). A perpendicular out-of-plane magnetic field is applied ($B_{Focus}$) which causes the holes to follow spin dependent cyclotron orbits, indicated by the red and blue lines in Figure \ref{fig:FocussingDiagram}d). When the diameter of a cyclotron orbit matches the distance between the injector and collector QPC, charge builds up in the collector and a peak in the collector voltage (and hence $R_{Focus}$) is observed. In the absence of spin-orbit interaction the magnetic focussing peaks occur at magnetic fields given by \cite{van_houten_coherent_1989}
\begin{equation}\label{eq:ClassicalFocussing}
    B_{Focus} = \frac{2 \hbar k_F}{e d_{Focus}}
\end{equation}
where $k_F$ is the Fermi momentum and $d_{Focus}$ is the distance between injector and collector QPC (focussing diameter). In the presence of a Rashba SOI, the first focussing peak splits due to the spin-dependent cyclotron orbits in the 2D region.\cite{rokhinson_spin_2004, zulicke_magnetic_2007, lee_influence_2021}
\section{\label{sec:TG}Top gate dependence of focussing peak splitting}
\begin{figure}
    \centering
    \includegraphics[width = 0.35\textwidth]{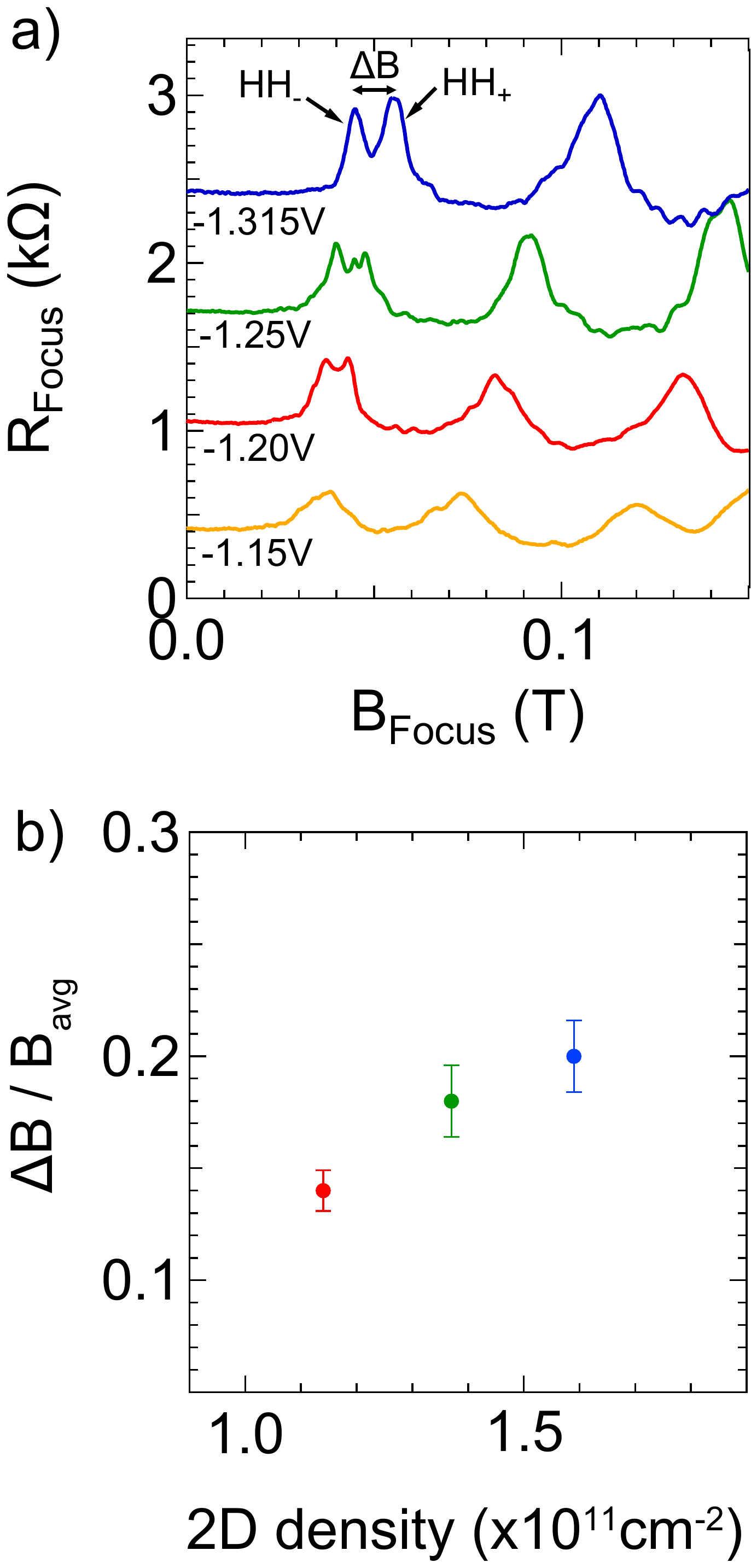}
    \caption{\textbf{a)} Focussing for different $V_{TG}$. $d_{Focus} = 2300nm$. More negative $V_{TG}$ results in a larger $n_{2D}$ and stronger Rashba SOI, increasing the splitting of the first focussing peak. \textbf{b)} Splitting between the magnetic focussing peaks as a function of $n_{2D}$. Error bars are the uncertainty in the peak position of a double gaussian fit.}
    \label{fig:VtgFocussing}
\end{figure}
In this section we use the voltage applied to the overall top gate of our sample to change the strength of the Rashba SOI and hence focussing peak splitting. As discussed in Section \ref{sec:HolesVsElectrons}, the Rashba spin-orbit term in 2D hole systems is given by
\begin{equation}
    \mathcal{H}_{R} \propto \beta \frac{E_z}{\Delta_{HH-LH}}k^3
\end{equation}
where $E_z$ is the electric field across the 2D interface, $\Delta_{HH-LH}$ is the splitting between the heavy hole (HH) and light hole (LH) levels in the quantum well and $k$ is the Fermi momentum. Increasing $\lvert V_{TG} \rvert$ increases both $E_z$ and $k_F$, while only causing a slight increase in $\Delta_{HH-LH}$ due to the quantum well confinement. The net result is an increase in the magnitude of $\mathcal{H}_R$ as $\lvert V_{TG} \rvert$ is increased, which causes an increase in the splitting of the first focussing peak. 

Figure \ref{fig:VtgFocussing}a) shows focussing measurements for four different values of $V_{TG}$. $V_{TG}$ was first set to $-1.15V$, corresponding to a density of $n_{2D} = 0.83\times 10^{11} cm^{-2}$ (bottom trace - yellow). Multiple evenly spaced focussing peaks can be observed, indicating low scattering and specular reflections from the gates.\cite{van_houten_coherent_1989} At $V_{TG} = -1.15V$ the Rashba SOI is not yet large enough to resolve a spin split first focussing peak. As $V_{TG}$ is made more negative, $n_{2D}$ increases and we observe multiple  effects on the focussing peaks. First, the focussing peaks move to higher $B_{Focus}$ since $k_F$ increases (Equation \ref{eq:ClassicalFocussing}). Second, there is an increase in the amplitude of the focussing peaks. This is due to an increase of the hole velocity, which causes a reduction in scattering as the holes travel through the 2D region from injector to detector.

The third effect of increasing $V_{TG}$ is to increase the Rashba splitting of the HH states, causing the first focussing peak to split. At $V_{TG}$ = -1.20V (red trace) the first focussing peak develops into a double peak, as the spin splitting is now able to be resolved. At $V_{TG}$ = -1.25V (green trace) the splitting of the first focussing peak is larger (with some additional structure due to branching flow and interference effects).\cite{van_houten_coherent_1989, aidala_imaging_2007, bladwell_interference_2017, lee_influence_2021} At the most negative $V_{TG} = -1.315V$ (top trace - blue) the spin split first peak can be completely resolved. This increase in splitting is qualitatively as expected, as a more negative $V_{TG}$ increases the Rashba SOI.

We now quantify the splitting of the first focussing peak. To measure the peak splitting, a double Gaussian is fit to the first focussing peak doublet to find the peak spacing ($\Delta B$). Since the magnitude of $\Delta B$ will depend on $d_{Focus}$ (see Equation \ref{eq:ClassicalFocussing}), the dependence on $d_{focus}$ can be removed by dividing by the central peak location ($B_{avg}$). Figure \ref{fig:VtgFocussing}b) shows the normalised splitting ($\Delta B/B_{avg}$) as a function of $n_{2D}$. An increase in the peak splitting is seen as $V_{TG}$ is made more negative (larger $n_{2D}$), with $\Delta B/B_{avg}$ approaching 0.2 (i.e. the splitting is approximately 20\% of $\hbar \omega_c$). The lowest density is not included in Figure \ref{fig:VtgFocussing}b) as a splitting of the first peak cannot be resolved.

These results show that it is possible to tune the spatial separation of spins in a 2D hole system using only a voltage applied to a top gate. In addition, the separation can be tuned over a wide range within the same sample - from too small to resolve (at $V_{TG} = -1.15V$) to complete separation (at $V_{TG} = -1.315V$). This tuneable spin separation makes 2D hole systems ideal for studying the spatial separation of spin states.

\section{\label{sec:Comparison}Comparing focussing to bandstructure calculations and Shubnikov-de Haas oscillations}
\begin{figure*}
    \centering
    \includegraphics[width = \textwidth]{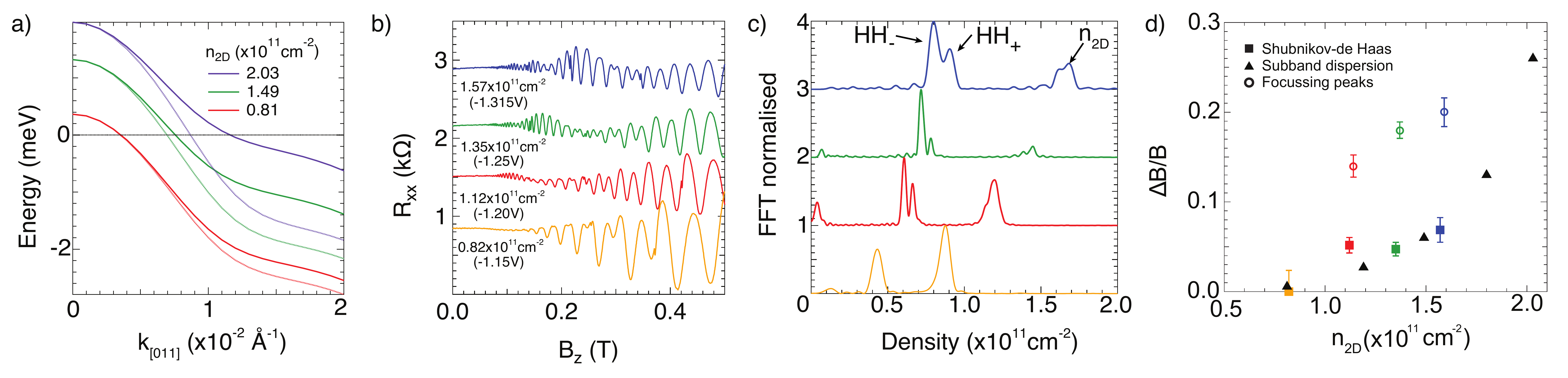}
    \caption{Comparing focussing peak splitting to Nextnano subband dispersion and Shubnikov-de Haas oscillations. \textbf{a)} Dispersion of the HH subbands at different $n_{2D}$ from a 6x6 k.p nextnano calculation. Solid lines indicate the HH+ subband, while dashed lines indicate HH-. \textbf{b)} Shubnikov-de Haas oscillations obtained from the focussing sample. \textbf{c)} FFT of the oscillations in a). Peaks correspond to the density of each spin chirality (HH+ and HH-) and the total density ($n_{2D}$). \textbf{d)} Comparison of the predicted focussing peak splitting from Shubnikov-de Haas (solid squares) and Nextnano (solid triangles) to the measured focussing peak splitting (open circles).}
    \label{fig:ShubsFocussing}
\end{figure*}
Next we compare the splitting of the first magnetic focussing peak to a predicted splitting from other methods. We use two methods to predict the focussing peak splitting - bandstructure calculations (using Nextnano) and Shubnikov-de Haas measurements.

Figure \ref{fig:ShubsFocussing}a) shows the calculated HH subband from Nextnano over the density range of our sample. As $n_{2D}$ is increased (more negative $V_{TG}$) the HH subband splits into two spin chiralities (HH+ and HH-). From the values of $k_+$ and $k_-$ at the Fermi energy ($E_F$ = 0) it is possible to predict a focussing peak splitting ($\Delta B/B_{avg}$). Equation~\ref{eq:ClassicalFocussing} shows that $\Delta k/k_F = \Delta B/B_{avg}$ where $\Delta k = k_{HH+} - k_{HH-}$. (This approximation is valid provided $\Delta k/k_F << 1$).

Shubnikov-de Haas oscillations can also be used to predict a focussing peak splitting. In the absence of a spin-orbit interaction, Shubnikov-de Haas oscillations are periodic in $1/B$, with a frequency proportional to $n$ ($f = n h/e$). The addition of a spin-orbit interaction creates two Fermi surfaces, each with a different spin chirality. The Fermi surfaces have different areas, resulting in two distinct frequencies in Shubnikov-de Haas oscillations\cite{stormer_energy_1983, papadakis_effect_1999} which can be used to predict the splitting of magnetic focussing peaks.

Figure \ref{fig:ShubsFocussing}b) shows Shubnikov-de Haas oscillations measured at the same $V_{TG}$ (same $n_{2D}$) as the focussing measurement on the same sample. To predict a focussing peak splitting ($\Delta B/B_{avg}$) the density of each spin chirality is found from an FFT of the Shubnikov-de Haas oscillations (Figure \ref{fig:ShubsFocussing}c). These densities are then used to find the momentum of each chirality ($k = \sqrt{4\pi n}$). $\Delta B$ can then be found by substituting $k$ into Equation \ref{eq:ClassicalFocussing}. Similarly, $B_{avg}$ can be found by substituting $k_F = \sqrt{2\pi n_{2D}}$ into Equation \ref{eq:ClassicalFocussing}.

Finally, we compare the measured focussing peak splitting to a predcited splitting from $k\cdot p$ calculations and Shubnikov-de Haas oscillations in Figure \ref{fig:ShubsFocussing}d). There is good agreement between Shubnikov-de Haas measurements (coloured squares) and Nextnano calculations (black triangles). However, the splitting of the magnetic focussing peaks (open circles) is consistently larger than the value predicted by Nextnano or Shubnikov-de Haas measurements. This suggests that the focussing peak splitting does not only depend the magnitude of the Rashba SOI.

To ensure that the split gates on the focussing sample did not affect the 2D measurement, Shubnikov-de Haas oscillations were also measured on a hall bar fabricated on the same wafer (W713). These values have good agreement with the Shubnikov-de Haas measurements on the focussing sample, demonstrating that the split gates on the focussing sample have a negligible impact on the Shubnikov-de Haas measurements. This shows that the split gates are not the cause of the disagreement between the predicted and measured focussing peak splittings.

Some additional explanations for the larger than expected splitting of the magnetic focussing peaks can also be ruled out. Comparing the focussing peak splitting with a 2D measurement (Shubnikov-de Haas oscillations) rules out any 2D effects such as Fermi surface distortion as this should also affect Shubnikov-de Haas measurements. Effects from lateral biasing of the QPC split gates\cite{lo_controlled_2017} can also be ruled out, as all measurements were performed with the split gates biased as symmetrically as possible to remain on the $G = 2e^2/h$ conductance plateau. Problems with the magnet used to create $B_{Focus}$ are also unlikely. The hysteresis of the magnet has been fully corrected and the magnitude of the magnetic field was verified using a Hall sensor on the sample probe. One possibility is additional complex spin dynamics created by the injector and detector QPCs. These dynamics could be created by the rapid spatial variation of the electrostatic potential in the injector and collector QPC, which may lead to non-adiabatic spin evolution.\cite{culcer_spin_2006} Additionally, recent theoretical work has found coupling between gate electric fields and heavy hole/light hole bands in GaAs.\cite{philippopoulos_pseudospin-electric_2020} This coupling may also be significant in the presence of a magnetic field, as in a focussing measurement and could account for the larger focussing peak spacing.
\section{\label{sec:Conclusions}Conclusions}
In this work we have investigated the 2D density dependence of magnetic focussing in hole systems with a Rashba spin-orbit interaction. We demonstrated control over the focussing peak splitting using the voltage applied to an overall top gate, showing that the splitting can be tuned from too small to resolve to a complete separation of the spin peaks. The magnitude of the peak splitting was compared to Shubnikov-de Haas oscillations and bandstructure calculations, and we found that the peak splitting is consistently larger than expected. This result may indicate the presence of complex spin dynamics or additional coupling between heavy hole and light hole bands.\cite{culcer_spin_2006, philippopoulos_pseudospin-electric_2020} The result suggests that care must be take when quantitatively relating the spacing of hole focussing peaks to the size of the Rashba spin-orbit interaction.
\section{Acknowledgements}
S.D. Liles and M.J. Rendell contributed equally to this work. The authors would like to thank S. Bladwell, O.P. Sushkov, D. Culcer and U. Z{\"u}elicke for many valuable discussions. Devices were fabricated at the UNSW node of the Australian National Fabrication Facility (ANFF). This research was funded (partially or fully) by the Australian Government through the Australian Research Council Discovery Project Scheme; and by the the UKRI Engineering and Physical Sciences Research Council [grant number CE170100039].
\bibliography{RashbaFocussingPaper}
\end{document}